# Bohr's Relational Holism and the classical-quantum Interaction[1]

Mauro Dorato
Department of Philosophy Communication and Media Studies
University of Rome "Roma Tre",
Via Ostiense 234, 00144, Rome, Italy
dorato@uniroma3.it

## 1 Introduction: a conflict in Bohr's philosophy?

Bohr's philosophy of quantum mechanics has often been charged for what is allegedly one of its major shortcomings, namely the advocacy of an unambiguous classical/quantum distinction (let me refer to this view with the label *the distinction thesis*). As is well known, such a distinction is needed to defend Bohr's view that any communicable measurement outcome must presuppose a classically describable instrument, with respect to which any reference to the quantum of action can be neglected (Bohr 1958, 4). Critics have then often insisted on the fact that the distinction in question is hopelessly vague (Bell 1987, 29) or at least strongly contextual (Ghirardi 2004), so that Bohr's interpretation of quantum mechanics suffers from the same vagueness and adhoc-ness. The resulting problem is, allegedly, a renunciation to describe the dynamical interaction between system and apparatus in a physically precise, theoretically based and non-contextual way, and therefore to offer a much-needed solution to the measurement problem.

In my paper I will present and critically discuss the main strategies that Bohr used and could have used to defend from this charge his interpretation of quantum mechanics. In particular, in the first part I will reassess the main arguments that Bohr used to advocate the indispensability of a classical framework to refer to quantum phenomena by trying to look at them from a new angle. I will then go on to discussing the nature of the indispensible *link* between classical measurement apparatuses and observed system that he also advocated. Typically, this link has been interpreted as a mere neopositivistic appeal to the fact that it is *meaningless* to talk about state-dependent properties of quantum entities independently of a measurement apparatus (Redhead 1987, 49-51, and Beller and Fine 1994). On the contrary, other authoritative scholars have rejected this minimalistic reading by stressing the fact that Bohr's view implies the presence of a holistic nonseparability between quantum system and classical apparatuses (Bohm 1951, Folse 1985, Faye 1991, Whitaker 2004, 1324). Howard (2004) and Tanona (2004, 691) explicitly interpret Bohr's relational view of a measurement by invoking the notion of *entanglement*. And it is clear that in this latter case, Bohr should have offered an explicit and well-articulated theory of measurement, a challenge that has been accepted among others also by Zinkernagel (2016) who, by relying on Landau and Litshitz's brief treatment (1981, 2-26), insisted that the quantum system interacts with only a *part* of a classical apparatus. In order to evaluate this discussion and give due emphasis to Bohr's holistic understanding of a quantum "phenomenon" (Bohr 1935 and 1958 among other sources), it is important to distinguish among different senses of Bohr's "holism" and consider which of these can be reconciled with "the distinction thesis" that he also explicitly advocated.

---

[1] Thanks to Henry Folse and Jan Faye for their attentive reading of a previous draft of the paper.

Let me add at the outset that in the following I will *not* try to argue that Bohr's interpretation of quantum mechanics is free from any conceptual difficulties. Nevertheless, I will try to correct some frequent misunderstandings that many of his critics have fallen prey to. After the 1935 confrontation with Einstein, it is well known that philosophers and physicists attributed to Bohr a definite victory over Einstein's criticism. But since the late sixties' surge of interest in the foundations of physics caused by Bell's theorem and his sympathy for alternative formulations of quantum mechanics (Bohmian mechanics and dynamical reduction models), Bohr has become been regarded as responsible – and not just by philosophers – for having "brainwashed a whole generation of physicists into thinking that the job was done 50 years ago" (Gell-Mann 1976, 29). It is about time to achieve a more balanced picture of Bohr's contribution to the philosophy of quantum mechanics. The conclusion that Bohr's interpretation of the formalism is untenable can *only* be established by giving his arguments as much force as possible, which is what I will try to do in the following by remaining as faithful as possible to his published work.

## 2 Bohr's recourse to classical concepts

Various misunderstandings of Bohr's philosophy of quantum mechanics have certainly been favored by the obscurities of his prose that he himself later acknowledged (Bohr 1949, 233),[2] as well as by overly polemical remarks due to some his brilliant opponents. Here is one example:

> Rather than being disturbed by the ambiguity in principle ... Bohr seemed to take satisfaction in it. He seemed to revel in contradictions, for example between "wave" and "particle," that seem to appear in any attempt to go beyond the pragmatic level. Not to resolve these contradictions and ambiguities, but rather to reconcile us to them, he put forward a philosophy, which he called "complementarity." (Bell 1987, 189)

A possible source of an important "contradiction" that to my knowledge has not received sufficient attention is given by the conflict between the sharp separation thesis and Bohr's

---

[2] "Rereading these passages, I am deeply aware of the inefficiency of expression which must have made it very difficult to appreciate the trend of the argumentation aiming to bring out the essential ambiguity involved in a reference to physical attributes of objects when dealing with phenomena where no sharp distinction can be made between the behaviour of the objects themselves and their interaction with the measuring instruments". Here he is referring to his Bohr (1935)'s reply to EPR.

holism, that is, his defense of the inseparability of the quantum systems from classical describable measurement apparatuses. As evidence for the former thesis, consider the following quotation: "The essentially new feature in the analysis of quantum phenomena is, however, the introduction of a fundamental distinction between the measuring apparatus and the objects under investigation." (1963, 3-4) Even though, as we will see, Bohr does *not* identify "the quantum" with "objects under investigation", it is well known that for him the "measuring apparatus" must be described with the language of classical physics, while the object of investigation is typically an atomic particle, not another classical object. Evidence for Bohr's full endorsement of the holistic nature of any measurement interaction (the latter thesis), is provided a little further down the same text: "While, within the scope of classical physics, the interaction between object and apparatus can be neglected or, if necessary, compensated for, in quantum physics this interaction thus forms an inseparable part of the phenomenon." (Bohr 1963, 4)

This problem is in part a side effect of a persisting lack of consensus on what his use of "classical physics" really amounts to. In order to make some progress on this question, it is important to distinguish among at least six related reasons that motivated Bohr's reliance on classical concepts. Even though some of these motivations have been already amply investigated, the perspective that I am stressing here is hopefully able to shed some additional light on each of them as well as on their logical relationship. My conciseness in this section of the paper can be justified in view of what comes next. Schematically, classical physics is needed

1. To defend the intersubjective validity of physical knowledge.

2. To provide a solid epistemic evidence for, and assign meaning to, claims about non-directly observable but *mind-independent*, real physical entities, that are codified in an otherwise abstract symbolism.

3. By using the Principle of Correspondence, to defend the continuity of quantum theory with

classical physics in a period characterized by radical scientific revolutions.

4. To be able to rely on a physical realm to which Heisenberg's indeterminacy Principle does *not* apply.

5. Following Einstein's distinction (Einstein 1919),[3] to formulate quantum mechanics as a *theory of principle* based on classical physics, and not as a constructive theory, in the same sense in which the special theory of relativity was formulated as a theory of principles (Bohr 1958, 5)[4]. This was meant to guarantee the completeness of the theory on the basis of the practical impossibility to use the wave function to describe classical apparatuses and atomic particles.

6. To replace states of superpositions between object and apparatus by mixed states in which both have classical, definite properties, to be specified as a function of the particular measurement setup (Howard 1994).

Let me expand in some more details on the five points above, with the exception of 6, which will be developed in the last section of the paper.

1. Bohr's insistence on the indispensability of classical physics is first and foremost justified by the need of communicating in a *non ambiguous way* the results of scientific research and therefore, in particular, of experiments. Such a non-ambiguity, on which Bohr insists so much, is guaranteed by our reliance on a *shared language*. This point is of crucial importance and it should not be downplayed.[5] As I understand it, "shared language" in Bohr is a technical term, since it refers to *plain, ordinary language, supplemented and enriched by physical concepts derived from classical physics* (Bohr 1958, 3). The role of a shared language is explicitly motivated by Bohr's vigorous and explicit refusal of a solipsistic or subjectivistic interpretation of quantum physics that we would lapse into if this theory implied some sort of relativity of experimental outcomes to the observation of *single* individuals or to mental observations of human beings in general.[6] In a word, given the indispensability of a shared language in the report of holistic quantum phenomena, the consequent elimination of any reference to subjective, mental states is blocked. The fact that language is shared necessarily leads to the view that a measurement is an "irreversible registration", as Wheeler often put it *a propos of* Bohr (Wheeler 1998, 337-338).[7]

---

[3] In referring to this article, I will be quoting from the 1954 reprint.

[4] This point has been already stressed by Bub (2000), but used to defend an antirealist philosophy of quantum mechanics.

[5] See Petersen (1968).

2. The necessity of relying on a shared language takes us to the strictly related, second point, *evidence*. This second point does not just express the trivial fact that any piece of evidence in favor of physical theories, quantum mechanics included, depends on empirical reports expressed by a shared language in technical sense above. More importantly, there are at least *two* remarks that need to be stressed:

2.1 The first relates to Faye's convincing suggestion that Bohr was an *entity realist* (Faye 1991).[8] The process of irreversible amplification due to the interaction between quantum systems and apparatuses described by a share language is in fact essentially *causal*. As a result of an inference to the best explanation, it follows that the macroscopic effects observed in any classically describable experimental device justify our belief in the reality of quantum entities regarded as causes of what we observe in the laboratories  (Hacking 1983).

2.2 In order to qualify what I stated in point 1 above, it is necessary to elaborate on what can be found in Bohr's work. His belief in the unrevisability of the description of the world afforded by the "shared language" *vis à vis* changing scientific worldviews does *not* entail that the new physical theories proposed in the 20[th] century do *not* put stress on the categories of ordinary language as they have been refined by classical physics (space, time, particle, wave etc.). It only entails that any imaginable evidential link between physical hypotheses and experiments presupposes that the manifest image of the world (Sellars 1963) – as it is expressed by our reliance on a shared language –cannot provide a *completely illusory* representation of the world around us. *If we had to accept this conclusion, we would have no reason to believe in physics to begin with*. In fact, our belief in a physical realm that goes beyond our unaided senses ultimately depends on evidence coming from our senses plus the referential power of "the shared language", which also entails our belief in the reality of causal relations.

---

[6] In some popular expositions we still read that quantum physics has vindicated the centrality of human beings in the physical world (due the role of the concept of observation).
[7] As is well known, Wheeler spent quite some time with Bohr in Copenhagen, and is therefore a reliable guide to report faithfully his thought.
[8] Faye's hypothesis that Bohr was an antirealist about quantum theory is more controversial since, given that independently of other readings of Bohr (Zinkernagel 2006), it is not clear whether entity realism can be defended *without* theory realism.

By using a distinction coming from an apparently unrelated philosophical corner, we could say that Bohr is not a *revisionist* philosopher of physics but rather a *descriptivist* one (for this important distinction, see Strawson 1959).[9] Strawson's main idea, that Bohr would have certainly endorsed, can be thus summarized: "there is a shared and universal conceptual scheme which we human beings have, and know that we have, and for which no *justification* in terms of more fundamental concepts or claims can be given." (Snowdon 2009, 32; my emphasis). On the contrary, *whenever there is some free room for different interpretations of physical theories*, revisionist philosophers favor worldview in which the main tenets of the manifest image sense are overthrown. An example of a revisionist metaphysics is configuration space realism (see papers in Ney and Albert 2013).[10]

The task of the philosopher of physics for Bohr is rather to describe the way in which the new quantum physics categorizes the world in interaction with the categories provided by the "shared language". To put it in a nutshell by using other words, Bohr is against "quantum fundamentalism" (Zinkernagel and Rugh 2011, Zinkernagel 2016), which, in my way of putting it, is the view that quantum physics grounds conceptually and epistemically – but not ontologically, given Bohr's entity realism – classical physics.

---

[9] The relevance for Bohr of an unrevisable conceptual scheme has been also stressed by MacKinnon (1982) and Murdoch (1985) and is of course at the basis of the Kantian readings of Bohr.
[10] Of course, whether one can go revisionist or not does not depend only on one's philosophical taste, but is almost always strongly contrained by physical theories.

3. In his search for the new quantum mechanics, Bohr wanted it to be the case that it implied some kind of limitation but not a complete overthrow of the classical concepts. Consequently, according to him quantum theory had to be regarded as a *rational generalization* of classical physics that applies to physical phenomena in which $h$ is non-negligible (Bokulich and Bokulich 2004, Faye 2014). In the same sense, general relativity is a generalization of Newtonian gravitation that applies when gravitational fields are strong, and special relativity is a generalization of Newtonian mechanics that applies when the relative velocity of bodies approaches $c$. Likewise, general relativity, with its manifolds with variable curvature generalizes special relativity, which is valid in infinitesimal regions of a Riemannian manifold, which can be considered to all effects Euclidean and therefore flat.

It is sometimes argued that Bohr's philosophy of quantum mechanics was mainly inspired by Einstein's operationalist analysis of the concept of simultaneity. This claim, if correct, would make Bohr into a full-blown neopositivist (see below). While the influence of Einstein's verificationist analysis on the physicists of Bohr's generation and on Bohr can hardly be overstated, reference to the 1905 paper in discussions about the epistemic status of quantum mechanics is much more frequent in Heisenberg and in Born (1971). Despite their different views about the theory, Bohr agreed wholeheartedly with Einstein's methodological approach to the construal of a new physical theory, which insisted on the importance to recover the previous theory from the successive one: "No fairer destiny could be allotted to any physical theory, than that it should of itself point out the way to the introduction of a more comprehensive theory, in which it lives on as a limiting case" (Einstein 1920, 102). In our case, Einstein's "pointing out" refers to the idea that any physical theory has in itself the germs for a generalization: it is in this sense that classical physics was for both Einstein and Bohr an indispensible heuristic means. The validity and truth of classical physics, in all three of the above-mentioned cases, is only *restricted* to some values of the variables intervening in the physical laws and not to all. The fact that classical mechanics cannot be extended to all possible values, as was thought beforehand, for Bohr did not mean that classical physics had to be abandoned, since the very concepts of "generalization" and "limiting case" imply – against unrealistic philosophies of science based on incommensurability (Faye 2014) or pessimistic metainduction – a full commensurability between the quantum and the classical epistemic and ontic domain. In a word, the epistemic reliability of classical physics was the main justification for an extension to the quantum realm via the rational generalization principle.

Furthermore, "mixed treatments" of quantum domains with classical theories[11] that were fundamental in establishing the new mechanics (Bohr's 1913 model of the atom) already implied that the latter had to be regarded in any case as approximately true, otherwise it would have been heuristically useless in predicting new phenomena: "The problem with which physicists were confronted was therefore to develop a *rational generalization* of classical physics, which would permit the harmonious incorporation of the quantum of action" (Bohr 1963, 2, my emphasis). We should realize that the importance of correspondence rules was particularly felt in a period in which, given the revolutionary changes happening in physics, a principle of continuity had to be reaffirmed in order to guarantee the rationality of scientific progress.

---

[11] This is common phenomenon in the history of 20[th] century physics: semi-classical quantum gravity mixes *quantum* field theory with *classical* general relativity, and is regarded as an approximation to a full theory of quantum gravity.

4. The epistemic, foundational role of classical physics depends also on the fact that a classical measurement apparatus is not subject to Heisenberg's indeterminacy relations. Unlike atomic objects, a classical object has a well-defined position in spacetime and *at the same time* obeys the causal laws of classical mechanics (conservation of momentum). According to the distinction thesis, complementary magnitudes of the apparatus are always simultaneously sharp. However, as a consequence of Bohr's complementarity principle, the experimental arrangements suitable for locating an atomic object in space and time, and those that are fit for a determination of momentum-energy, are mutually exclusive. Consequently, we cannot use *the same apparatus* to measure complementary magnitudes. Clearly, if an apparatus did not have a precise position before and during measurement (as confirmed by our senses), we could not measure the precise location where an electron hits a photographic plate. At the same time, the fact that, as we can see, a position-measurement apparatus always has *also* a definite momentum, i.e., it does not move, is what makes the measurement of the position of the particles possible. On the other hand, if the diaphragm suspended with springs did not have a precise position and momentum before during and after a measurement, we could not use the law of conservation of momentum to calculate the momentum of the particle that goes through it. In the next section, we will see whether Bohr's response to Einstein's thought experiment during the Solvay conference is not a withdrawal of the distinction thesis, since Bohr applies Heisenberg's relations also to the macroscopic, classical screen suspended on springs, thereby *apparently* abandoning this thesis.[12] In fact this is an aspect of Bohr's holism.

5. In order to provide a deeper understanding of Bohr's reliance on classical physics, it is

---

[12] The localizability of macro-objects in spacetime is important for an additional reason, which is linked to the fact that space and time are *principia individuationis*: their localizability is sufficient to ascribe them an identity On the contrary, quantum particles are indistinguishable and come to possess an identity only when they are localized by a classical apparatus.

illuminating to recall Einstein's distinction between *principle* physical theories and *constructive* physical theories (Einstein 1954, 227-232) and suggest that Bohr conceived of quantum mechanics as a *principle* theory – as Einstein did with special relativity – and not as a *constructive* theory (see also Bub 2000). The distinction is best explained with Einstein's own words: "constructive theories attempt to build up a picture of the more complex phenomena out of the materials of a relatively simple formal scheme from which they start out" (Einstein 1954, 228). A constructive theory for Einstein is the kinetic theory of gases, which explains thermal, macroscopic processes in terms of underlying molecular motions. On the contrary: "..The elements which form their basis and starting-point [i.e., of principle theories] are not hypothetically constructed but empirically discovered ones, general characteristics of natural processes, principles that give rise to mathematically formulated criteria which the separate processes or the theoretical representations of them have to satisfy" (1954, 228). His example of a principle theory is *thermodynamics*, which "seeks by analytical means to deduce necessary conditions, which separate events have to satisfy, from the universally experienced fact that perpetual motion is impossible".

Einstein regarded special relativity as a principle theory based on two very general postulates, namely the principle of relativity and the independence of the speed of light from the motion of the source. These postulates act as meta-laws (Lange 2013a, 2013b), that is, as very general constraints that all separate processes ("separate events") governed by all lower-level physical laws must obey. As plausibly argued by Bub (2000), in Bohr's philosophy of quantum mechanics the two general principles are the quantum postulate and the principle of correspondence, whose role is analogous to Einstein's two postulates. *Qua* axioms of the theory, they exclude any attempt to further explain them by constructing them causally or dynamically, in the precise sense that Einstein gave to "constructions". Of course, the two postulates of quantum mechanics do *not* function as meta-laws in exactly the same sense as the principle of relativity. Yet, when they were formulated, they fixed structural constraints about how to calculate transition probabilities (Bub 2000, 78). Furthermore, as argued in a later paper written together with Clifton and Halvorson, by changing somewhat Bohr's two principles in order to introduce three information-theoretic axioms, it is possible to make the analogy with Einstein's "theory of theory" approach even stronger. These three principles in fact "constrain the law-like behavior of physical systems" (Clifton, Bub and Halvorson 2003).

Even though Bohr was very probably unaware of the London Times article (Einstein 1919) and of its later 1954 reprint, it is rather striking that – in order to justify his philosophy of quantum mechanics – he refers explicitly to Einstein's purely *kinematic* foundations of the special theory of relativity. In his 1949 piece written in honour of Einstein we read:

> Notwithstanding all differences between the physical problems which have given rise to the development of relativity theory and quantum theory, respectively, a comparison of purely logical aspects of relativistic and complementary argumentation reveals striking similarities *as regards the renunciation of the absolute significance of conventional physical attributes of objects*. Also, *the neglect of the atomic constitution of the measuring instruments themselves, in the account of actual experience, is equally characteristic of the applications of relativity and quantum theory.* (Bohr 1949, 236, my emphasis)

Notice that in the first part of the italicized quotation, Bohr is directing our attention to the historical process of relativisation of physical quantities that were previously regarded as absolute ("the renunciation of the absolute significance of conventional physical attributes of objects"). Exactly as in the special theory, spatial and temporal intervals "taken by themselves" become relative to an inertial frame, in quantum mechanics it is the possession of definite physical properties that becomes relative to experimental situations.[13]

---

[13] Rovelli's relational interpretation of quantum mechanics (1996) extends Bohr's approach to any physical system, thereby avoiding any reference to a classical apparatus. For an evaluation of Rovelli's attempt, see Dorato (2016).

The second part of Bohr's passage pursues the analogy with relativity one step forward, since it makes explicit the non-constructive character of special relativity. Unlike what happens in Lorentz's *dynamical* theory of the inner constitution of the rulers, contractions of classical objects and temporal dilations of clocks don't need to be explained bottom-up from a constructive theory, since they can be "explained" geometrically from the structural constraints of a four-dimensional spacetime that codifies the two axioms of the theory. Contractions don't need to be explained from a dynamic theory because they are non-invariant (Dorato and Felline 2010, Dorato 2014).

In the same spirit, according to Bohr the two axioms/principles from which quantum mechanics depend (quantum postulate and the principle of complementarity) render futile any attempt to *construct* classical objects out of underlying quantum entities with their laws. If classical objects could be explained bottom up by a theory T that fully reduced them to quantum properties, we would lack any evidence for T to begin with, since evidence must ultimately be expressed in classical language for the reasons already given.

There are however some differences between the epistemic and methodological foundations of two theories of which Bohr was certainly aware: quantum physics is committed to a greater amount of "relationality" and holism than the special theory of relativity. In the latter theory there is an *invariant* that plays the crucial role: the four-dimensional metric with its light cone structure supersedes completely the relational nature of space and time "separately considered" (Minkowski 1952, 75).

In Bohr's philosophy of physics, the relativistic invariant metric might be matched by the indispensability of a description provided by classical physics (the "experimental invariant"). This analogy, however, does not extend very far: the necessary presence of a classical experimental setup does not avoid complementary descriptions: apparatuses capable of revealing the wave-like feature of an atomic object are incompatible with setups that are necessary to reveal its particle-like property. Consequently, the relational nature of a quantum property is by all means much more relevant than a classical one. Despite the fact that any experiment takes place in a particular frame, the fact that an observer O in frame F measures length L in relativity holds for all possible observers.

In any case, the principle/constructive distinction has a double function. On the one hand, it illustrates in a clear way that Bohr drew inspiration from Einstein's early methodological attitude. On the other hand, it is also very useful to account for their later divergence about the completeness of quantum mechanics, a fact that Bub (2000) and other scholars have neglected. In 1905 Einstein is *provisionally* construing special relativity as a principle theory because of the poorly understood, dual nature of the underlying electromagnetic components of rulers and rigid bodies (Brown 2005, 72-73). He was convinced that in physics one should first proceed with principle theories as the case of thermodynamics and only later attempt to hypothesize constructive theories, which provide a deeper understanding of physical phenomena (Howard 2015, section 6). To the extent that Einstein regarded quantum mechanics as a theory of principle that unified a lot of empirical phenomena without offering a deeper comprehension,[14] we can understand his distance from Bohr, who thought that no deeper explanation was forthcoming because no explanation was needed.

In fairness to Bohr, the fact that he regarded quantum mechanics as a principle theory for him did not entail the impossibility of any bottom-up explanation of the properties of classical bodies: "Although, of course, the existence of the quantum of action is ultimately responsible for the properties of the materials of which the measuring instruments are built and on which the functioning of the recording devices depends, this circumstance is *not relevant* for the problems of the adequacy and completeness of the quantum−mechanical description in its aspects here discussed." (Bohr 1949, 223, my emphasis). Against Zinkernagel (2015), we should note that passages like this seem to imply that the quantum realm is ontically more fundamental than the classical realm, especially in view of the fact that Bohr was an entity realist. Of course, as noted above, an ontic primacy intended in this sense is compatible with the epistemic primacy of the classical physics illustrated above.

---

[14] Not surprisingly, Bub (2000) ignores this aspect of Einstein's attitude toward quantum mechanics as principle theory, because he is mainly interested in using it to defend his informational approach to quantum mechanics (Bub 1997).

The irrelevance in question is based on a very *general* fact concerning any kind of measurement in physics: the need for approximations and idealization, that in the case of quantum mechanics is due to the scale of phenomena:

> …the smallness of the quantum of action compared with the actions involved in usual experience, including the arranging and handling of physical apparatus, is as essential in atomic physics as is the enormous number of atoms composing the world in the general theory of relativity which, as often pointed out, demands that dimensions of apparatus for measuring angles can be made small compared with the radius of curvature of space. (Bohr 1949, 238).

In addition to what we have presented in this section, we might add that the scale of the phenomena to be measured that renders a classical description unavoidable.

## 3 Bohr's relational holism and the classical-quantum interaction

We can now return to the question that I introduced at the beginning of the paper: is there a conflict between the distinction thesis and the holistic character with which Bohr characterized any quantum phenomenon? On the one hand, the previous section has presented the main reasons that drove Bohr to rely on the domain of classical physics as something *ontologically* distinct from that of quantum mechanics but necessary to gain empirical evidence about atomic objects. As his critics have insisted, however, this thesis, which for his approach is absolutely central, presupposes as a necessary condition the existence of a clear-cut, non-contextual criterion to distinguish the quantum from the classical realms.[15] On the other hand, however, he insists on the "*impossibility of any sharp separation between the behavior of atomic objects and the interaction with the measuring instruments which serve to define the conditions under which the phenomena appear*" (Bohr 1949, 210, emphasis in the original).

A defender of Bohr's interpretation could remark that the contradiction here is only apparent, as it is based on an ambiguity that is merely semantical: "lack of sharp separation" is compatible with "distinctness". When Bohr talks about the "impossibility of a sharp *separation*" he is referring to a relational link between quantum objects and classical apparatus, or even more radically, to a form of holism or non-separability that has been empirically proved after Bell's inequalities, and that EPR interpreted as a type of nonlocality that had to be avoided. Without some further assumption, the friends of Bohr insist, this notion of "non-sharp separation" is compatible with Bohr's assumption that the classical and the quantum domains possess radically incompatible properties and are therefore *ontically distinct*.

However, Bohr's enemies note, where and how can we draw a clear-cut, non-ambiguous distinction between the two ontological levels if in the *physical* process of measurement these two levels *interact*? Suppose that the inseparability of the parts of a holistic "phenomenon" in Bohr's sense implies also for Bohr an entanglement of the atomic objects with a classical apparatus (see Howard 1994, 2004). If one assumes at the same time the linearity of the evolution equation and the *completeness* of the theory, as Bohr certainly did, one has then the problem of explaining the observed definiteness of the outcomes that are revealed by classical apparatuses. To avoid this problem, Zinkernagel follows Landau and Lifshitz's approach to the measurement problem (1981) by (1) denying the existence of an entanglement relation between the atomic entity and the *whole* classical apparatus, and by (2) claiming that the former interacts only with a *part* of the latter (Zinkernagel 2016, 11). However, the problem represents itself also with this suggestion: how are we to distinguish the "part" of a classical object from the whole of it? How big must the part be for it to follow the laws of classical mechanics? Is there a precise, theory-derived criterion that enables us to answer these questions? At this point Bohr's foes could insist that the claim that the distinction between part of a classical object and the whole is contextual counts as a purely pragmatic solution to the problem. As such, isn't it a just a way to sweep the dust under the carpet?

---

[15] Furthermore, how can we come to know the wave function of the universe if nothing classical can interact with it from the outside?

In order to answer these important questions, we need to dig deeper into the meaning of "non-separable" in Bohr's account of a "phenomenon", and consequently into his theory of measurement, no matter how weak it is. In fact, without such a clarification, it is not clear whether Bohr's "relational holism" is really compatible with the existence – on which his interpretation depends – of two clearly distinct descriptions of the world, one referring to the quantum and the other to the classical realm.

First of all, all interpreters seem to agree that by talking about a "*finite, uncontrollable interaction*" between system and apparatus, Bohr assigns to the interaction a *physical* meaning. The interaction may *cause* an increase of information, but in virtue of this very statement one has to assume that here we are dealing with a physical, non-fictitious cause (i.e., the interaction). There is ample evidence for this claim in his texts: in his famous reply to EPR's paper, for example, he discusses the consequences of measuring precisely the position of an atomic particle with a double-slit diaphragm and a photographical plate bolted to a common support.[16] The interaction of the electron with the slit is *finite* because of the discrete, indivisible nature of the quantum of action. The fact that it is *uncontrollable* means that in these experimental conditions it impossible to come to know the momentum of the electron because it cannot be *manipulated* at will: the result of this impossibility is unpredictability and indeterminism. While Bohr is not always explicit about the nature of this uncontrollability, it is a consequence of a practical impossibility due to the smallness of the atomic particles, as well as to the fact that their dynamical behavior is analogous to the jump of an electron from an orbit to the other when hit by photons: in his response piece in the Schilpp volume, he refers in this respect to spontaneous emission processes and radioactive decays, attributing these examples to Einstein's theory of radiation published in 1917 (Bohr 1949, 202). All of this terminology shows that the measurement interaction for Bohr is a *physical*, *irreversible* process.

If this is to be taken for granted, there is still some important disagreement as to whether this interaction makes room for a genuine, even though qualified, "collapse" of the wave function (Zinkernagel 2016), or it rather only causes an increase of our information about the system (Faye 1991). In the latter case, the wave function is merely an instrumental, book keeping device, devoid of any ontological significance, while in the former the fact that the wave function symbolizes something, stands for something "out there", needs to be further discussed.

Bohr's *contextual* view of measurement does not help *per se* to solve this dispute. It is in fact also non-controversial that the particular complementary property of an atomic entity revealed by an experiment *depends* on the particular setting that is appropriate to measure that property (either spatiotemporal properties or dynamical properties). However, there seem to be at least four senses of dependence that need to disentangled, since they are logically independent:

---

[16] The fact that this is an ideal experiment, as Bohr recognizes, is irrelevant for the point he wants to make.

beyond the problem of establishing which of these senses (maybe more than one) is defended by Bohr, I am interested in asking which of them renders his whole philosophical approach least amenable to criticism.

The first sense of dependence is also the weakest, since it is compatible with assuming that also *before* a measurement any quantum system possesses a definite value of both position and momentum. These values however are unknowable in principle, due to the uncontrollable *disturbance* that the macroscopic apparatus causes in the measured system. Whether Bohr defended this position before 1935 is still controversial, but after EPR's he certainly rejected it (see also Faye 2014, 13).[17] After all, talking of "disturbance" implies that there is a definite property that existed before the measurement and that it is disturbed by it (Beller and Fine 1994), or at least it does not exclude it. It is not plausible to believe that Bohr did not see this logical possibility.

In other words, this first sense of dependence is weak because it holds also in classical physics: the result of an experiment on a physical system – the experimental answer – depends on the kind of question we ask and is in any case influenced by the apparatus. However, in the classical case it is *in principle* possible to discount the effect of the apparatus on the measured system, so that it is much more reasonable to assume that experiments reveal pre-existing values of any physical magnitudes (even if it might be *de facto* rather difficult to determine them). As is well known, according to Bohr's Complementarity Principle the story is indeed quite different. Bohr never tires of insisting on the fact that, as a consequence of Heisenberg' principle of indeterminacy and the mathematical formalism, asking a "position-question" to an electron, and therefore providing a *spatiotemporal description*, precludes in principle asking a momentum-question, and therefore giving a *causal description* of the process, and conversely. The two experimental settings are mutually exclusive, even though for an exhaustive description of an atomic object we need both properties.

---

[17] By referring to EPR, in his 1949 Bohr wrote "there is in a case like that just considered no question of a mechanical disturbance of the system under investigation during the last critical stage of the measuring procedure (Bohr 1949, 235). The "disturbance argument" was suggested by Heisenberg in 1930, in his optical, "microscope argument".

The second, stronger sense of dependence, which is nearer to Bohr's position after 1935, suggests that there is *no* pre-existing value that the apparatus measures and the coming into existence of a definite value is due to the measurement interaction. This is the sense of measurement that is assumed by EPR-type of arguments: since it is not possible that a measurement in a region of space causes an instantaneous coming into existence of the value of the distant particle, quantum theory is incomplete. Interestingly not only does John Bell – the staunchest opponent of Bohr – attribute him this second sense of "dependence", but he even agrees with him that this the right view to take about measurement in quantum mechanics: " the word [measurement] very strongly suggest the ascertaining of some pre-existing property…Quantum experiments are not just like that, *as we learnt especially from Bohr*. The results have to be regarded as the joint product of "system" and "apparatus", the complete experimental set-up" (quoted in Whitaker 1989, 180, my emphasis).

An obvious objection to the distinction between the two senses might come from neopositivist corners. For a neopositivist, in fact, there is no difference between the claim that there are definite magnitudes before measurement that cannot be measured due to a "disturbance" (first sense), and the claim that it is the measurement that causes the acquisition of a definite position or momentum previously undetermined (second sense). In both cases it would be meaningless to talk about magnitudes without a measurement, and in any case one cannot measure simultaneously position and momentum. Interestingly, *to the extent that* in Bohr's approach to quantum measurement there has been a transition from the first sense of dependence of measurement results from apparatuses to the second, there would be an additional reason not to count him as a neopositivist. Since without a verificationist approach one cannot move from the impossibility of simultaneous measurement to the impossibility of a common value, the fact that took so seriously EPR's argument is another piece of evidence that he was not a neopositivist Conversely, as remarked as Whitaker (2004), the impossibility of common values conjoined with the projection postulate implies the impossibility of measuring complementary magnitudes.[18]

Given this understanding of "dependence on measurement", how did Bohr conceive of an atomic entity? If it were exclusively a particle or a wave, we could not explain the experiments: in the former case, it would not generate interference patterns; in the latter, it would not localize in the photographic plate. Likewise, for Bohr it could not be both a particle *and* a wave in the sense of De-Broglie's pilot wave theory, where the particles' velocities are fixed by guiding waves (Bacciagaluppi and Valentini, 28). More plausibly, for him it is neither a particle nor a wave until it interacts with a particular apparatus. One way to cash this out would be to argue that an atomic particle has *two dispositional properties* to be both a particle and a wave, depending on "the experimental question we ask", that is, in the dispositionalist language, on the particular *stimulus*. Certainly this position seems closer to Heisenberg's thought (1958) but it is interesting to go over the reason why Bohr did not consider it, independently of well-known personal enmity with the German physicist.

We saw that two properties are complementary if and only if they are *mutually exclusive*

---

[18] "Once one has agreed to accept the [projection] postulate, it is easy to close the gap and to move from the impossibility of common values to that of simultaneous measurement. A measurement of momentum should leave the system in an eigenfunction of momentum; a measurement of position should leave it in an eigenfunction of position. Since there can be no common eigenfunctions, there can be no simultaneous measurements" (Whitaker 2004, 1313)

and *jointly exhaustive* (see Murdoch 1987). Kinematical and dynamical properties of atomic entities are *mutually exclusive* because they cannot be attributed to the same system at the same time via a single measurement, but they are also jointly exhaustive because in classical terms, they are both necessary for a complete knowledge of the dynamical properties of the entity.

However, from a *dispositionalistic* point of view, before the experiment we can attribute the *same* particle *two irreducible and intrinsic dispositions* (powers) for manifesting a particle-like behavior *and* for a wave-like behavior, depending on the experiment. When one disposition is inhibited, the other is selected as a result of the interaction with the chosen experimental setup, in such a way that they two can never be manifest at the same time in a single experiment. The presence of two dispositional properties is the reason why complementary descriptions about spatiotemporal trajectories and causal properties of the atomic particle cannot be used simultaneously. Analogously, for the trajectory of the particle (its position) and the interference pattern: if we know through which slit the particle enters, the wave-like disposition is inhibited by the experimental setup. Conversely if we observe the manifestation of the wave-like disposition (the interference effect), we cannot know where the particle went because the particle-like disposition is inhibited. In both cases, the possession of both dispositional properties before the measurement interaction does not imply the possession of definite values for the respective magnitudes. If the possession of definite values is regarded as the categorical basis of the dispositional properties, the two dispositional properties are to be regarded, as they should, as irreducible to the categorical basis. In a word, an electron is neither a particle nor a wave, but the relevant dispositional properties coexist in the same particle.[19] *

In order to make the reasons for Bohr's refusal of this interpretation explicit, it is sufficient to point out that the holistic link between the atomic particle and the apparatus stressed by Bohr involves the manifestation of the dispositions – *the measurement event* – and not the two dispositions *before* measurement. In this sense, the dispositional approach – if it is not a merely verbal redescription of the holistic interaction between system and apparatus[20] – would not help to clarify the nature of this interaction. Possibly, this approach might make this link even less perspicuous, by suggesting that an introduction of dispositions that get selected by the classical apparatus can by itself solve the measurement problem (as suggested by Suárez 2004).

---

[19] One could adopt the same approach by invoking the existence of just one dispositional property. I am not going to explore this alternative here.

The third, more controversial sense of dependence calls into play the "nonseparability" or "indivisibility" of the measurement apparatus from the measured entity. It is important to remark at the outset that the fact that before measurement the measured system has no definite value (second sense of dependence) *per se* is compatible with the fact that before measurement quantum entities and classical instruments have an independent reality: the coming into existence of the definiteness of the conjugate variable depends on the apparatus, but the latter has properties that don't depend on the former. In other words, according to the second sense, the dependence is asymmetric.

This remark provides evidence for the fact that if indeed Bohr understood the "impossibility of any sharp separation between the behavior of atomic objects and the interaction with the measuring instruments which serve to define the conditions under which the phenomena appear" (Bohr 1949, 210) as a form of holism, he was referring to a third sense of dependence, which implies the other two but is not implied by them. Likewise, in 1927, at the Como lecture, he wrote "the quantum postulate implies that any observation of atomic phenomena will involve an interaction with the agency of observation not to be neglected. Accordingly, an independent reality in the ordinary physical sense can neither be ascribed to the phenomena nor to the agencies of observation" (quoted in Zinkernagel 2016, 11). In a passage like this, it is clear that the dependence we are trying to clarify is *symmetric*: not only does the reality of the atomic phenomenon depend on the apparatus, but also the converse is the case. Before trying to understand the nature of this third sense of dependence in Bohr's philosophy, we should ask whether it is reasonable to attribute it to Bohr.

---

[20] In Dorato (2007) I argue that it is not.

In some cases, admittedly Bohr's texts allow multiple interpretations: "quantum phenomena find logical expression in the circumstance that any attempt at a well-defined subdivision [between particle and apparatus] would require a change in the experimental arrangement that precludes the appearance of the phenomenon itself." (Bohr 1956, 168). Here the change may be regarded as compatible with an asymmetric form of dependence, and therefore with the first two senses of dependence. This is probably why the neopositivist reading has it that according to Bohr it is meaningless to attribute a definite property to a quantum system unless we specify a particular measurement context, which functions as a frame of reference in the special theory of relativity: "To regard Bohr as endorsing a nonlocal or nonseparable conception of reality strains his carefully tailored language of measurement and his picture of the operational presuppositions on physical magnitudes posed by conditions of measurement" (Beller and Fine 2004, 26).

This neopositivist interpretation neglects the fact that the need to have a "well-defined subdivision" that Bohr is referring to in the quotation above *presupposes* that there is something to be divided, which in its turn implies that instruments and atomic particle form a one, a whole. This whole, however, cannot be studied without introducing some separation, whether conventional or not: given the holistic nature of what Bohr calls a phenomenon, "one must therefore cut out a partial system somewhere from the world, and one must make 'statements' or 'observations' just about this partial system" (Bohr 1985, 141).

Let us then take for granted that in Bohr's writings, ambiguous as they sometimes may be, there are important hints toward a stronger sense of holistic dependence such that, due to the measurement interaction, not only does the atomic particle lack a definite property, but also a part or the whole of the classical instrument does, because the description we give of it depends on the whole experimental condition. When we measure the exact momentum of an incoming particle by using a mobile screen suspended with springs (see Bohr 1949), we include *this* classical apparatus in the description of the holistic phenomenon in Bohr's sense. The fact that the mobile spring is subject to Heisenberg's relation implies that *it is treated as a quantum system*, independently of its size. In fact, as Howard has pointed out, the distinction between quantum and classical does not coincide with that between atomic system and apparatus. The fact that a measuring instrument can be described quantum mechanically implies that only some of its properties can be regarded as classical (that is, those that are correlated, e.g., the momentum of the particle before measurement and the momentum of the spring after measurement, see Howard 1994, 214), but this implies, in virtue of Heisenberg's uncertainty relations, that the position of the particle before and that of the mobile spring after measurement are entangled, because they both lack definite properties. For these reasons a measurement of the particle position in this particular situation is impossible.

Of course, talking of a third sense of dependence in terms of nonseparability or entanglement generates the problem of explaining our observations of a definite world within a theory that relies on a linear equation and is thought to be complete. However, since for Bohr there is clearly *no* room for a collapse of the wave function regarded as a physical process causally describable in spacetime, how do we justify the transition from a dynamic indivisibility between system and apparatus to the possession of a definite quantity on the part of the former?

Here I will focus only on three explanatory strategies. The first is to explain the transition from the quantum to the classical as the passage from an entangled pure state to a mixture (Howard 1994, 203),). The second appeals to a view of measurement regarded as an irreversible physical process, due to Landau and Lifshitz (1981) and revived among others by Zinkernagel (2016), and, the third invokes the Bohmian's holism of the undivided universe.

1) As acknowledged by Howard himself, his thought-provoking suggestion to regard Bohr's holism as a form of entanglement between the two systems is not grounded in his texts. Independently of questions of hermeneutic faithfulness, which here can be in part neglected, according to Bohr the third sense of dependence (the measurement-induced nonseparability between a property of the atomic system and that of the classical apparatus) is never such that the two systems are jointly in a pure state. For Bohr there is *no* measurement problem exactly because according to him there are no situation in which *both* properties of the system and the apparatuses before, during and after measurement are indeterminate, despite their symmetric dependence and despite the fact for Bohr a classical object *can* be attributed a quantum state (as in the case of the screen suspended with movable springs).

There are at least two reasons for this claim. The first has been explored in various ways in the previous section: to summarize, in order to confer some meaning to our experimental practice, we must assume that at the end of the measurement interactions there is a classical object endowed with well definite classical properties. The second reason stresses the idea that quantum mechanics is a theory of principle linked to some intrinsic, insurmountable limitations. If there existed an underlying causal description that we're able to construct or explain the definiteness of measurement outcomes bottom-up, this description in any case could not be given in a spatiotemporal language. To the extent that our form of understanding is limited to the intuitions of phenomena in space and time as Kant had it, this description would not add to our comprehension of quantum theory. From this conclusion, however, the tension between Bohr's need of a condition of sharp distinction between the classical and the quantum and his belief in the non-separate reality of atomic system and apparatuses – a nonseparability that does not prohibit treating a classical object as obeying Heisenberg's relations and therefore to quantum mechanics – has not been completely removed.

2. The second approach claims that Bohr's holism is to be interpreted as an entanglement between particles and, crucially, only *a part* of the classical, measuring apparatus (Zinkernagel 2016). The entanglement of the particle with a part of the classical apparatus calls into play Bohr's appeal to the necessity of a "cut" between entangled stuff: "one must therefore cut out a partial system somewhere from the world, and one must make 'statements' or 'observations' just about this partial system" (Bohr 1985, 141).

Zinkernagel draws inspiration from passages like these and, as we will see, from Landau and Lifshitz's treatment (1981). He claims that the evolution of the wave function describes a real process in the world that cannot be given a purely epistemic meaning. At the same time, however, it cannot be subject to more fundamental explanations in terms of a constructive theory: "We can … say that, for Bohr, the collapse is not physical in the sense of a physical wave (or something else) collapsing at a point. But it is a description – in fact the best or most complete description of something happening, namely the formation of a measurement record". (2004,14). In other words, not only is the wave function an abstract object defined in the mathematical model, but it also refers to an irreversible process of amplification in the physical world. This real process is the outcome of the entanglement of the atomic system with the part of the classical apparatus that is involved in producing a well-defined record. This interpretation of the measurement interaction can at least in part be traced back to Bohr, with the obvious proviso that the term "entanglement" does not explicitly occur in his work. Measurement results are due to "suitable amplifications devices with irreversible functioning such as, for example, permanent marks on a photographic plate" (Bohr 1954, 73).[21]

However, this solution, despite its ingenuity, raises some difficulties: (a) given the existence of an entanglement, doesn't Bohr need some sort of collapse theory, even if limited to the dynamic interaction between some part of the apparatus and the atomic systems? And if he does not need such a theory, as Zinkernagel claims by following faithfully Bohr's texts, (b) doesn't he still need a more precise account for the obtaining of a definite record in the part of the classic system that is entangled with the atomic particle? If these questions were left unanswered, typical objections concerning how to draw the boundary or a distinction between "a part" (which can enter in an entangled state) and a whole of a classical object (which cannot) would stick up their ugly heads again. It is at this stage that the notions of approximation and idealization might play an essential role.[22]

In order to rebut the well-known charge that Bohr does not justify in a non contextual way the cut between the classical and the quantum it seems necessary to exploit the importance of position in any form of measurement that has been often signaled also by Bell: the classical limit involves the obtaining of a well-defined spatial trajectory in a bubble chamber, or as mentioned in the above quotation, "the formation of permanent marks". The answer to the question: "how do we identify the part that is entangled with the system?" may then be provided by an appeal to an *approximation* of a quantum system with a classical system. So classicality is given by the existence of a well-defined trajectory. This is essentially Landau and Lifshitz's 1981 "Bohrian" account of measurement, which begins by noting that in order to describe the trajectories of atomic particles and attribute them dynamical properties, one needs apparatuses that can be regarded as classical "to a sufficient degree of accuracy" (Landau and Lifshitz 1981, 2). The two Russian physicists note that such apparatuses need not be macroscopic, but the deliberate vagueness of "sufficient degree of accuracy" allows us to count as apparatuses also microscopic instruments, which, I take it, is the source of Zinkernagel's appeal to the entanglement between particles and "parts of a classical object". In our case, the formula "sufficient degree of accuracy" refers to the track left by ionized molecules in a Wilson chamber; as noted by Landau and Lifshitz, the track is classical to the extent that it is very large with respect to the electrons, so that the latter becomes a quasi-classical object. On the other hand, they continue, the track measures the electron path with a very low degree of accuracy, which is the necessary price to pay for having definite outcomes and well-defined paths. In conclusion, the readings of the apparatus are described by quasi-classical functions and the fact that the apparatus is classical means that "all of these possible values have always definite value" (1981, 22).

---

[21] The irreversibility called into play here is rather different from the temporal evolution in decoherent approaches to quantum mechanics, given that the latter involve, as noted by Schlosshauer and Kamilleri, improper mixtures, and therefore quantum states (Schlosshauer and Kamilleri 2012, 5), while in Bohr's account the irreversibility concerns classical states.
[22] The notion of approximation or idealization is indeed crucial in Bohr's philosophy (Tanona 2004).

In order to comment on these neglected pages – to which, however, Zinkernagel makes explicit reference [23]– we should note that measures that are realized with a "sufficient degree of accuracy" are used throughout physics and not just in the context of quantum measurements. *The degree of accuracy cannot but depend on the aim at hand*, and it is for this reason that the boundary between classical and quantum must depend on the context. According to Bohr, this is compatible with the fact that, *for any given experimental context, the distinction between classical and quantum is always sharp and non-arbitrary.*[24] *Contextualism would imply vagueness only if we tried to distinguish the classical from the quantum across all experimental*

---

[23] I must thank Nino Zanghì for his suggestion to look at Landau's treatment as one of the best defenses of Bohr's interpretation.
[24] For Bohr, even though not for Heisenberg, the sharp distinction thesis did not imply a movable shift between the classical and the quantum. For him, there is a distinction between the contextuality of the boundary classical/quantum and its "changeability". He accepted the former but rejected the latter, as recorded by Heisenberg in a letter he wrote to Heelan (see Schlosshauer and Kamilleri 2012, 4).

*contexts.*

These italicized sentences go back to the main issue raised in this paper: a non-arbitrary, non-movable, distinction between classical and quantum is needed to cut definite outcomes out of an entangled mess (holism). On the other hand, contextualism and holism are called into play to take into account the principle of complementarity and the variability of the experimental setup. This alleged conflict is precisely the Achilles' heel that his enemies have been focused on. Is Bohr's contextualism really leading to a pernicious vagueness?

The conceptual situation that we are facing in this case is very similar to the famous sorite paradox. Even though the concept of being bald is vague– in certain cases we don't know how to distinguish between a bald person and a hairy one, and if pluck one hair at a time from a hairy person, there is no moment at which she becomes bald – at the *extremes* of the spectrum there is still a clear-cut distinction between baldness and hairiness. Likewise in the quantum/classical case: this fact, for Bohr is enough. If in any experiment we can find non-arbitrary borders at the extreme of the spectrum – say in terms of spatiotemporal paths – the distinction remains sharp and the vagueness of the borders in "intermediate" cases can be avoided.

In conclusion, whether the opposition between distinction and contextuality implies a violation of criteria of physical precision, as Bell has it, is of course dependent on one's own general philosophical intuitions of physics, or one's *stances* (van Fraassen 2002). As such, as of now the problem cannot be settled only on the basis of philosophical or conceptual arguments. Until, of course, theoretical and experimental progress forces us to revise these stances. However, this experimental progress so far is lacking, and the alternative theories are not producing new predictions. At best, non-standard interpretations or theories are based only on a superior explanatory power. This epistemic virtue, while so far not sufficient to consider them superior also on empirical grounds, is certainly sufficient to encourage their practitioners to further pursue new avenues: without Einstein's *philosophical* worry on no locality, we would not have quantum computing and quantum cryptography.